\documentclass{ws-p8-50x6-00}

\begin{document}

\title{Inelastic Hadron Diffraction\\ 
in High Energy Elastic Scattering of Nuclei}

\author{A. Ma{\L}ecki and D. Goc-Jag{\L}o}

\address{KEN Pedagogical University, Krak\'ow, Poland\\ 
E-mail: amalecki@ultra.ap.krakow.pl}

\author{M. Pallotta}

\address{Laboratori Nazionali di Frascati dell'INFN, Italy\\
E-mail: ma.pallotta@tiscalinet.it}  


\maketitle

\abstracts{
The r\^{o}le of inelastic diffraction in elastic scattering of nuclei is
studied in the formalism of \emph{diffractive limit}. The results obtained
for scattering of the $\alpha$--particles on light nuclei show that the
nucleonic diffraction is especially important at large momentum transfers
where the Glauber model of geometric diffraction fails. }

\section{Introduction}
Inelastic hadron diffraction refers to quasi-elastic transitions in which 
the colliding particles conserve their dominant quantum numbers. High energy 
of collision facilitates to satisfy the condition of coherence thus
assuring that interaction does not change the character of the particles.

High energy elastic scattering of hadrons is thus generally built of:\\
i) elastic diffraction which results from a strong absorption
connected to an opening of a variety of inelastic channels and by analogy
to classical optics it can be described in terms of geometrical shapes of
colliding particles;\\
ii) inelastic diffraction which appears as a quantum phenomenon
related to intrinsic structure of hadrons and has no classical analogy.

It will be shown that elastic (classical or geometrical) diffraction is 
\mbox{dominating} at low momentum transfers, especially in the region of
the forward peak. Instead, the inelastic (quantum or dynamical) diffraction is 
particularly important at high momentum transfers.

\section{Diffractive transitions}

The most convenient basis for calculating the diffractive amplitude is 
provided by the experimentally revealed division of inelastic channels
into 
the diffractive and non--diffractive transitions. This means that the 
space of physical states is decomposable (with respect to the initial state) 
into subspaces of diffractive $[D]$ and non--diffractive states $[\sim D]$. In
other words there exists a unitary operator $U$ which is reducible
(block--diagonal) 
in the Hilbert space:
\bea
\langle k|U|j\rangle=0\;\;\;\;\;\mbox{for any}\;\;\; 
|j\rangle\in[D]\;\;\;\mbox{and}\;\;\;|k\rangle\in[\sim D].
\eea

Expanding the initial $|i\rangle$ and final $|f\rangle$ states in the
basis of $|Uj\rangle$ states one obtains the transition amplitude:
\bea
T_{fi}\equiv\langle
f|T|i\rangle=\sum_{|j\rangle,|k\rangle\in[D]}U_{fk}t_{kj}U_{ij}^{*}
\eea
where $U_{fk}=\langle f|Uk\rangle,\;\;\;t_{kj}=\langle
k|U^{\dagger}TU|j\rangle\;\;\;\mbox{and}\;\;\;
U_{ij}=\langle i|Uj\rangle.$

In terms of the normal operator $\Lambda\equiv \widehat{1}-U$ this
reads:
\bea
T_{fi}=t_{ii}\delta_{fi}-\sum_{|k\rangle}\Lambda_{fk}t_{ki}-\sum_{|j\rangle}
t_{fj}\Lambda_{ij}^{*}-\sum_{|j\rangle,|k\rangle}\Lambda_{fk}t_{kj}\Lambda_{ki}^{*}
\nonumber \\
=t_{i}\delta_{fi}-N_{fi}(T_{0})\Lambda_{fi}t_{i}-
t_{f}\Lambda_{if}^{*}N_{if}^{*}(T_{0}^{\dagger})\;+
\sum_{|j\rangle\in[D]}N_{fj}(T_{0})\Lambda_{fj}t_{j}\Lambda_{ij}^{*}
\eea
where $t_{j}=t_{jj}$ are the diagonal matrix elements of
$T_{0}\equiv U^{\dagger}TU$
and the undimensional quantities $N_{kj}$ are defined as 
\bea
N_{kj}(T_{0})=\frac{1}{\Lambda_{kj}t_{jj}}\sum_{|l\rangle\in[D]}
\Lambda_{kl}t_{lj}.
\eea

If the subspace $[D]$ contains a huge number of diffractive states then
$N_{kj}=N\to\infty$ for any pair of states $|k\rangle$ and
$|j\rangle$. In fact, since
$\Lambda$ is a non--singular operator its matrix
elements change smoothly under the change of diffractive states. This
leads to a considerable simplification of Eq.(3):
\bea
T_{fi}=t_{i}\delta_{fi}-N\Big(\Lambda_{fi}t_{i}+t_{f}\Lambda_{if}^{*}-
\sum_{|j\rangle\in{D}}\Lambda_{fj}t_{j}\Lambda_{ij}^{*}\Big).
\eea

In general, the effect of non-diagonal transitions inside the diffractive
subspace $[D]$ gets factorized. In the case of elastic scattering
one has:
\bea
T_{ii}=t_{i}+N\sum_{|j\rangle\in[D]}|\Lambda_{ij}|^{2}\left(t_{j}-t_{i}\right)=
t_{i}+g_{i}N\left(t_{av}-t_{i}\right)
\eea
where $g_{i}=\sum_{|j\rangle}|\Lambda_{ij}|^{2}=2Re\Lambda_{ii}$ and 
$t_{av}=\sum_{|j\rangle}|\Lambda_{ij}|^{2}t_{j}/g_{i}$
is the average of the diagonal matrix elements $t_j$.
The expressions of the form $N\Delta t$ where $\Delta t$ represents
diversity of $t_j$ over the subspace of diffractive states $[D]$ are to be
considered in the double \emph{diffractive limit} [1]: $N\to\infty,\;
\Delta t\to 0 $ such that $N\Delta t$ is finite. The
contribution of inelastic diffraction is thus built as an 
\emph{infinite sum} of the \emph{infinitesimal} contributions from all
possible intermediate states belonging to $[D]$.

\section{Phenomenology of diffractons}

Our numerical analysis of elastic scattering was done in the framework of
a model where the diffractive states are built of a two--hadron core
(representing the ground state) and some quanta describing diffractive
excitations:
\bea
|j\rangle=|i\rangle+\;|n;\vec{b}_{1}\;.\;.\;.\vec{b}_{n}\rangle.
\eea
The configurations of these quasi--particles [2] (called
\emph{diffractons}) are specified by a
number of constituents and their impact parameters. Thus 
\bea
\frac{1}{g_{i}}\sum_{|j\rangle\in[D]}|\Lambda_{ij}|^{2}\ldots=
\sum_{n=1}^\infty P_{n}\int d^{2}b_{1}\ldots d^{2}b_{n}\prod_{k=1}^{n}
|\Psi(b_{k})|^{2}\ldots
\eea 
where $|\Psi(b_{k})|^{2}$ describe a spatial distribution of
diffractons (with respect to the core) and $P_{n}$ are probabilities of
their number.

The diagonal elements of $T_{0}$ are assumed to be purely absorptive. In
terms of the real profile functions we have:
\bea
N(t_{j}-t_{i})=i(1-\Gamma_{0})\lim_{N\to\infty, \gamma\to 0} 
N\sum_{k=1}^{n}\gamma\big(\vec{b}-\vec{b}_{k}\big)
\eea
with $t_{i}=i\Gamma_{0}$ representing the hadronic core and $\gamma$'s
corresponding to diffractons. 
The diffracton model thus explicitly accounts for the
\emph{geometrical} diffraction on an absorbing hadronic bulk and the
\emph{dynamical} diffraction corresponding to intermediate transitions
between diffractive states.

For elementary hadron collisions the elastic scattering profile resulting
from Eqs. (6), (8) and (9) would read:
$\Gamma_{el}(b)=\Gamma_{0}+(1-\Gamma_{0})\Gamma_{n}$. In the case of
nucleus--nucleus scattering there are, howewer, two sources of dynamical
diffraction. At the nuclear level the diffractive excitations correspond 
to various configurations of nucleons while the partonic compositness of
nucleons gives rise to the subnuclear diffraction. Therefore the
elastic profile reads:
\bea
\Gamma_{el}(b)=\Gamma_{0}+(1-\Gamma_{0})\Gamma_{N}+(1-\Gamma_{0})
(1-\Gamma_{N})\Gamma_{n}
\eea    
where $\Gamma_{N}$ and $\Gamma_{n}$ refer to nuclear and nucleonic
diffractiveness, respectively.

In our calculations we took the geometrical diffraction profile of two
nuclei with $A$ and $B$ nucleons as $\Gamma_{0}=(1-\gamma_{AB})^{AB}$
which corresponds to multiple scattering between nucleons frozen
in the nuclear ground states. The function $\gamma_{AB}$ and the dynamical
profiles $\Gamma_{N}$ and $\Gamma_{n}$ were taken in the Gaussian form: 
$(\sigma/4\pi R^{2})\exp(-b^{2}/2R^{2})$. The parameters of $\gamma_{AB}$
are related to the total nucleon--nucleon cross--section and the radii of
the colliding nuclei. The parameters of the profiles $\Gamma_{N}$ and
$\Gamma_{n}$ were determined from fitting the elastic differential 
cross--section based on Eq.(10) to experimental data.  

\begin{figure} [t]
\begin{center}
\epsfxsize=9.47cm 
\epsfbox{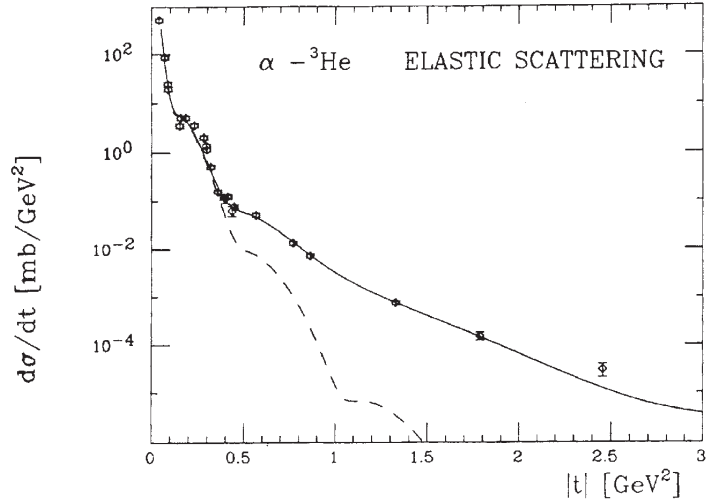} 
\caption{The $^{4}$He -- $^{3}$He elastic cross--section vs.
the squared momentum transfer.}
\end{center}
\end{figure}

In the presented fit to the $\alpha$--$^{3}$He data at $p_{\alpha}=7$
GeV/c [3] we found $R_{N}=0.90$ fm, $\sigma_{N}=77$ mb, $R_{n}=0.29$ fm
and $\sigma_{n}=84$ mb. The comparison with the contribution of the
geometrical profile $\Gamma_{0}$ alone (dashed curve) shows that the
dynamical terms $\Gamma_{N}$ and $\Gamma_{n}$ become important at
high momentum transfers where the Glauber model of multiple scattering
fails.

Our analysis reveals the existence of three interaction radii in
nucleus--nucleus scattering: the external radius $R_{AB}$ describes
geometrical diffraction on a black disc, the larger intrinsic radius
$R_{N}$ characterises the dynamics of interacting nucleons while the
smaller radius $R_{n}$ may be interpreted as the size of partonic clusters
inside the nucleons. However, the large value of subnuclear cross-section 
$\sigma_{n}>\sigma_{N}$ could signal an enhanced population of partons
and their deconfinement in the whole nuclear interior.

\end{document}